\documentclass[showemail,amsfonts,pre]{revtex4} 
\usepackage{amsmath}
\usepackage{amssymb}
\usepackage{graphicx}
\usepackage{bm}
\begin{document}

\title{Network Growth with Preferential Attachment for High Indegree and Low Outdegree}
\author{Volkan Sevim$^{1}$}
\email{sevim@scs.fsu.edu}
\author{Per Arne Rikvold$^{1,2}$}
\email{rikvold@scs.fsu.edu}

\affiliation{$^{1}$ School of Computational Science, Center for Materials Research
and Technology, and Department of Physics, Florida State University,
Tallahassee, FL 32306-4120, USA\\
 $^{2}$ National High Magnetic Field Laboratory, Tallahassee, FL
32310-3706, USA}

\date{\today}

\begin{abstract}
We study the growth of a directed transportation network, such as a food web,
in which links carry resources. We propose a growth process in which 
new nodes (or species) preferentially attach to
existing nodes with high indegree (in food-web language, number of prey) 
and low outdegree (or number of predators). This scheme, which we call {\it 
inverse preferential attachment\/}, is intended to maximize the amount of 
resources available to each new node. 
We show that the outdegree (predator) distribution decays at least 
exponentially fast for large outdegree and is continuously tunable 
between an exponential distribution and a delta function. 
The indegree (prey) distribution is poissonian in the large-network limit.
\end{abstract}


\maketitle
\newcommand{\kout}{k_i} \newcommand{\kin}{k'_i} \newcommand{\nk}{n_k^*}
\newcommand{\newzeta}{m/z_m^*} \newcommand{\zstar}{z_m^*}

\newcommand{\nkin}{n_{k'}^*}

\section{Introduction}
Directed networks that transport a resource, such as energy, from one or 
several sources to a large number of consumers are important in many 
areas of science \cite{Banavar:1999}. 
Among such networks, food webs provide an example of great interest, both 
from a purely scientific point of view, and because of their importance for 
nature-conservation efforts \cite{Rooney2006}. 
Although knowledge is rapidly accumulating about the {\it structure\/} 
of food webs 
\cite{Rooney2006,CommunityFoodWebs,Camacho:2002B,Camacho:2002A,Stouffer:2005,Camacho:2006}, 
and static models have been developed to describe some of their 
statistical properties 
\cite{CommunityFoodWebs,Camacho:2002B,Camacho:2002A,Martinez:2000}, 
we are as yet only beginning to develop an understanding of 
the processes by which such networks are formed and evolve under the 
influence of speciation, invasion, and extinction of 
interacting species 
\cite{DROS01B,DROS04,Quince2005,Rossberg:2006,Rikvold:2007,Rikvold:2007b}. 

An important aspect of the network structure of food webs is that they have
degree distributions that generally decay quite fast with increasing degree 
-- in most cases at least exponentially 
\cite{Camacho:2002B,Camacho:2002A,Martinez:2000,Rikvold:2007}. 
This is in sharp contrast to the class of networks known as scale-free, 
which have power-law degree distributions 
\cite{Barabasi:1999,Barabasi:2002}. While there has been a veritable 
explosion of research on scale-free networks, there has been no similar 
surge of interest in networks with rapidly convergent degree distributions.
Most food webs, and some (but not all \cite{Guimera:2004}) 
 transportation networks, such as the North American power
grid \cite{Albert:2004},  and the European railway network \cite{Kurant:2006} belong to this class.
Much work remains to be done before a comprehensive understanding 
of the mechanisms by which such networks evolve is reached. 

As a step toward the development of such an understanding, we here 
propose a network growth scheme that produces
a poissonian indegree distribution (in food-web language: prey distribution) 
and an outdegree (predator) distribution that is continuously tunable 
between an exponential distribution and a delta function. We note that these degree
distributions do not agree with current food-web theory. In particular, the indegree distribution
produced by the niche model \cite{Martinez:2000} has an exponential tail \cite{Camacho:2002B}. 
It has been claimed that models with an exponentially decaying probability
of preying on a given fraction of species with lower or equal niche values are 
capable of producing food webs that are structurally in agreement with empirical data \cite{Stouffer:2005}. However, it is not clear why this condition is necessary or why schemes 
that invoke no physical mechanisms (as in the niche model) are able to produce such webs.
Therefore, other plausible schemes should also be explored.

Our model employs a scheme in which new nodes (species) 
attach to existing nodes with a preference for nodes $i$
with high indegree $k'_i$ and low outdegree $k_i$. In food-web terms, 
this corresponds to a prospective predator choosing prey that have a large 
number of resources (represented by the large indegree), while the
competition from previously established predators should be as small as 
possible (low outdegree). Among the influences on the network growth process 
mentioned above (speciation, invasion, and extinction), we have thus chosen 
to focus on invasion and/or speciation. By ignoring extinction, we 
essentially model the early phase of steady network growth. 
The proposed growth process corresponds to a probability of attachment, 
\begin{equation}
\Pi(k'_{i},k_{i})\propto(k'_{i}/k_{i})^{\gamma}
\label{eq:one}
\end{equation}
with $\gamma \ge 0$. 
This attachment scheme is the direct opposite of the ``rich get richer" 
scheme of preferential attachment with an 
attachment probability proportional to the total degree $l_i = k'_i+k_i$, 
which is known to produce scale-free networks and has been 
studied in a myriad of 
variations over the last decade \cite{Barabasi:1999,Barabasi:2002}. 
To emphasize this difference, we shall call the scheme proposed here 
{\it inverse preferential attachment\/}. 
[We note that nonlinear forms of the ``rich get richer" scheme 
with a probability of
attachment proportional to $l_{i}^{\alpha}$ with $\alpha>0$
have also been studied. However, except for
$\alpha=1$, no $\alpha>0$ leads to a power-law degree distribution
\cite{Krapivsky:2000}.]

In a previous paper \cite{Sevim:2006}, we studied a simplified
version of the scheme proposed here, 
in which a new node makes a constant number
of incoming links ($\kin=\mathrm{const}.)$, and $\gamma=1$. In this
simplified version, the probability of attachment, $\Pi(k_{i})\propto1/k_{i}$,
depends only on the outdegree (the number of predators). We
calculated the outdegree distribution for this simplified model both
analytically and by Monte Carlo simulations. It is given by the self-consistent
equation
\begin{equation}
n_{k}^{*}=(k+1)(\newzeta)^{k}\frac{\Gamma\big(1+\newzeta\big)}
{\Gamma\big(k+2+\newzeta\big)} \;,
\label{eq:nk}
\end{equation}
where
\begin{equation}
z_{m}^{*}=\sum_{j=0}^{\infty}{\frac{n_{j}^{*}}{j+1}}
\;,
\label{eq:zstar}
\end{equation}
and $\Gamma(x)$ represents the Gamma function.

\section{Model and results}
In the present paper, we investigate the general form of the attachment 
probability presented in Eq.~(\ref{eq:one}).
With this form we relax both of the restrictions of the simplified 
model: we do not fix the indegree (number of
prey) for the new nodes, so that each can make a different number
of links, and we also vary the exponent $\gamma$. This makes full
analytical treatment much harder, and the results for the outdegree
distribution presented here are therefore only numerical. The indegree
distribution, however, is found analytically to be a poissonian in
the large-network limit.

The generalized attachment process proceeds as follows. We start the growth 
process with $N_{0}$ nodes and
assign each initial node an indegree $m\le N_{0}$. 
These nodes act as source nodes as they
are not connected to any other node at the beginning. Actually, the
attributes of the initial nodes have no significance for the statistics
because the final size of the network, $N_{\mathrm{max}}+N_{0}$,
is much larger than $N_{0}$. In each time step, we add a new isolated
node. Then, we give the new node $m$ chances to establish a link
to an existing node with probability
\begin{equation}
\Pi(\kin,\kout)=\frac{1}{Z}\left(\frac{\kin}{\kout+1}\right)^{\gamma}
\;,
\label{eq:attachprob}
\end{equation}
where
\begin{equation}
Z=\sum_{i=1}^{N}\left({\frac{\kin}{\kout+1}}\right)^{\gamma}
\label{eq:Z}
\end{equation}
with $\gamma \ge 0$. Here, $N$ denotes the number of existing nodes
at that time step. We use $\kout+1$ in the denominator to prevent
a divergence for $\kout=0$. Multiple links between two nodes are
not allowed. The direction of a link is from the old node to the new
one.

We implement the growth process in Monte Carlo simulations as follows.
We seed the system with $N_{0}$ source nodes, each with indegree
$m$, and introduce a new node in each Monte Carlo step. To create
links between the new node and the existing ones, we pick existing
nodes, $i,$ one by one and calculate the probabilities of attachment,
$\Pi(\kin,\kout)$. Then, we generate a random number, $r,$ and attach
the new node to node $i$ if $r<\Pi$. We repeat this procedure until
all existing nodes in the network are tested, i.e., till the sweep
is completed. Since $\sum_{i}\Pi(k'_{i},k_{i})=1$, a new node makes
on average one connection per sweep. We sweep the whole network $m$
times, so that $\langle\kin\rangle=\langle\kout\rangle=m$.
The new node
is kept in the system, even if it does not acquire any links. However,
a node with $k'=0$ stays isolated throughout the growth since the
probability of attachment to it is zero. We stop the growth
when the network size $N$ reaches $N_{\mathrm{max}}+N_{0}$ nodes
with $N_{\mathrm{max}}=10^{5}$. We average over fifteen independent
runs for each value of $m$ and $\gamma$.

\begin{figure}[t]
\centering
\includegraphics[clip,scale=0.3]{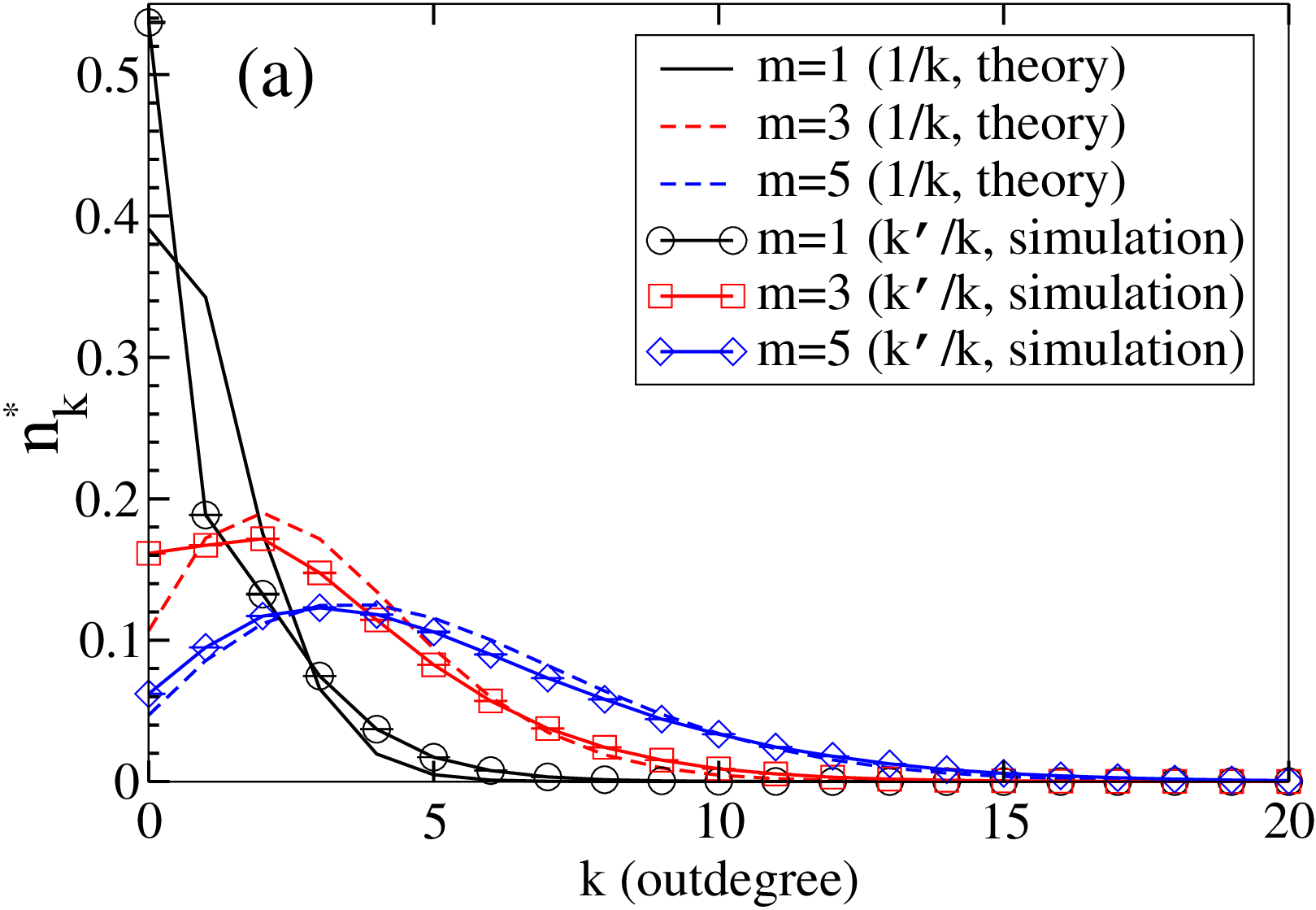}\\
\includegraphics[clip,scale=0.3]{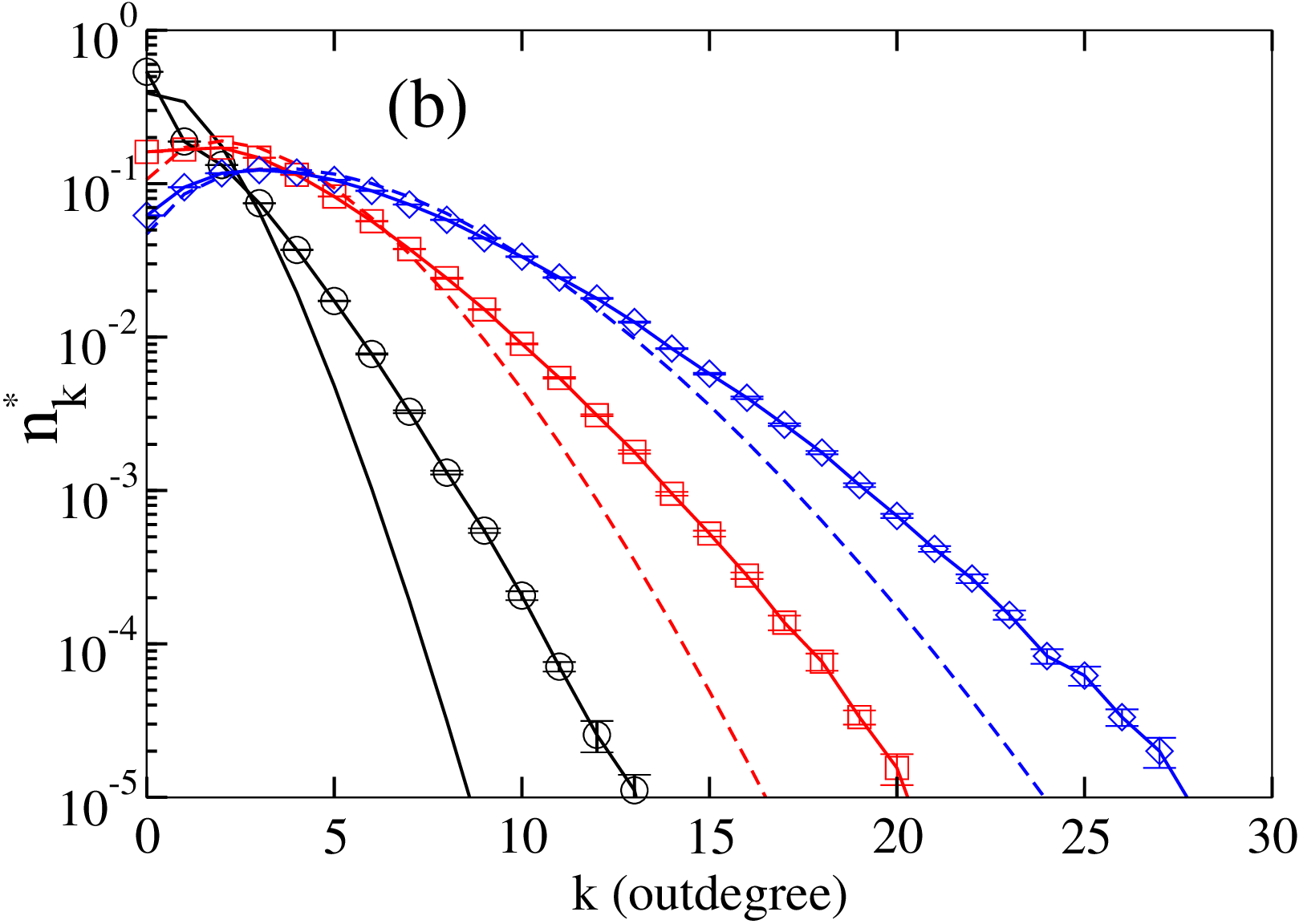} 
\caption{\label{fig:gamma1-out}Outdegree distributions for $\gamma=1$ and
$m=1,$ 3, and 5 with $N_{0}=10$ shown on linear (a) and log-linear
(b) scales. The simulations were stopped when the network size reached
$N_{0}+10^{5}$ nodes. Each curve (with symbols) represents an average
over fifteen runs. The curves without symbols are the theoretical
outdegree distributions for our simplified model with $\Pi(k)\propto1/k$.
Both the $k'/k$ and $1/k$ models yield the same distribution for
$m\gg1$. As $k$ is a discrete variable, lines connecting the symbols
are merely guides to the eye.}
\end{figure}

We first tested the case of $\gamma=1$ to compare the outdegree distribution
of the full $k'/k$ model to the outdegree distribution of our simplified
$1/k$ model, Eqs.~(\ref{eq:nk}-\ref{eq:zstar}), 
which, for large $k$, decays like $k\mu^{k}/\Gamma(k)$,
where $\mu$ is a constant. As seen in Fig.~\ref{fig:gamma1-out},
the outdegree distribution for the $k'/k$ model also
decays faster than exponentially for large $k$. However, the dependence 
on the variable indegree leads to a broadening of the outdegree 
distribution: decreased
probabilities for $k\approx m$, and compensating increased probabilities
for $k\gg m$ and $k\ll m$. In the limit of large $m$, the central part 
of the outdegree
distribution of the $k'/k$ model approaches that of the $1/k$ model.

\begin{figure}[t]
\centering
\includegraphics[scale=0.3]{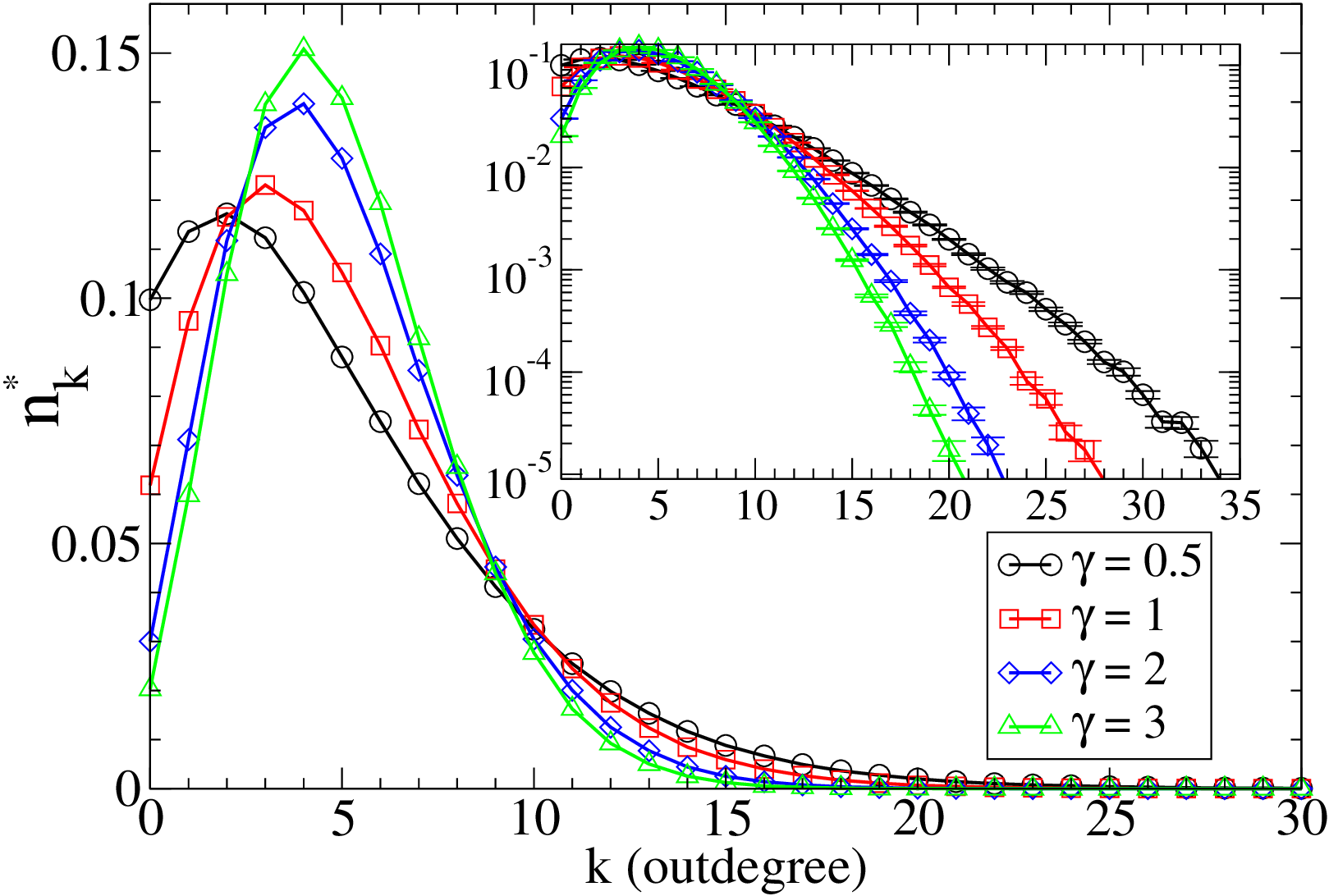} 
\caption{\label{fig:vargamma-out}Outdegree distributions for $\gamma=0.5,$
1, 2, and 3, with network size $N_{0}+10^{5}$ nodes, and $m=5.$
Each curve is averaged over fifteen runs. Inset: The same distributions
shown on a log-linear scale. The lines connecting the symbols are guides to the eye.}
\end{figure}
The outdegree distribution of the general model 
also varies with $\gamma.$ Higher values
of $\gamma$ sharpen the peak of the distribution around the mean
outdegree, $m$, as it increases the tendency of the new nodes to
prefer existing nodes with a higher value of $k'/k$ 
(Fig.~\ref{fig:vargamma-out}).
In the limit $\gamma\rightarrow\infty$ one should obtain a 
delta function at $k=m$. Similarly, lower values of $\gamma$
relax the constraint and flatten the outdegree distribution. The limiting
case, $\gamma=0$, corresponds to growth without preferential attachment,
which yields an exponential outdegree distribution of mean $m$
\cite{Barabasi:2002,Sevim:2006}. The lines connecting the symbols
are guides to the eye.
\begin{figure}[t]
\centering
\includegraphics[clip,scale=0.3]{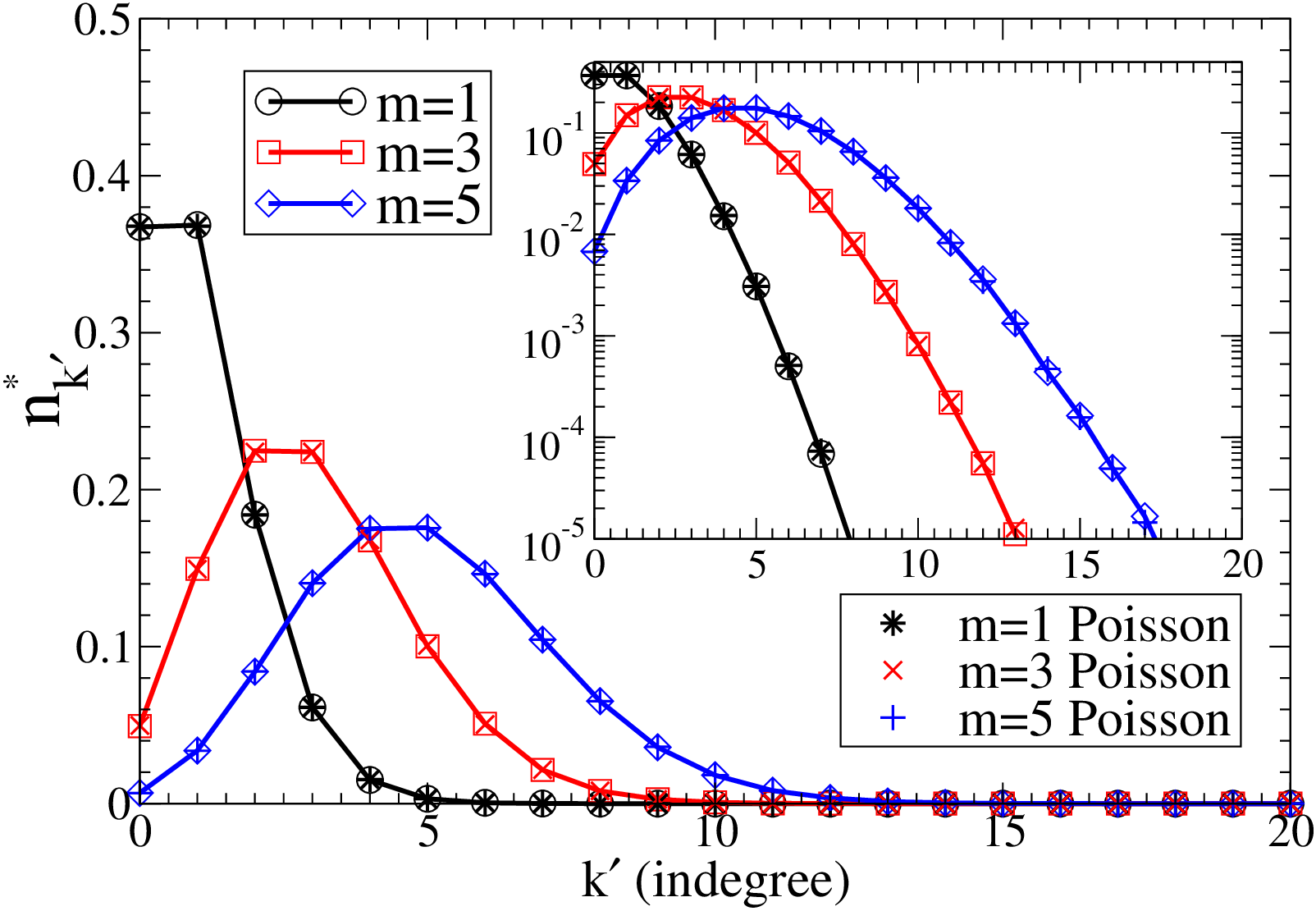} 
\caption{\label{fig:gamma1-in}Indegree distributions for $\gamma=1$ and
$m=1,$ 3, and 5 with $N_{0}=10$. The simulations were stopped when
the network size reached $N_{0}+10^{5}$ nodes. Each curve represents
an average over fifteen runs. The error bars are smaller than the
symbol sizes. Inset: The same distributions shown on a log-linear
scale. The symbols $\ast,\times,$ and + show the Poisson distribution,
Eq. \eqref{eq:poisson}, for $m=1,$ 3, and 5, respectively. 
The lines connecting the symbols are guides to the eye. See text
for details. }
\end{figure}

\begin{figure}[t]
\centering
\includegraphics[scale=0.3]{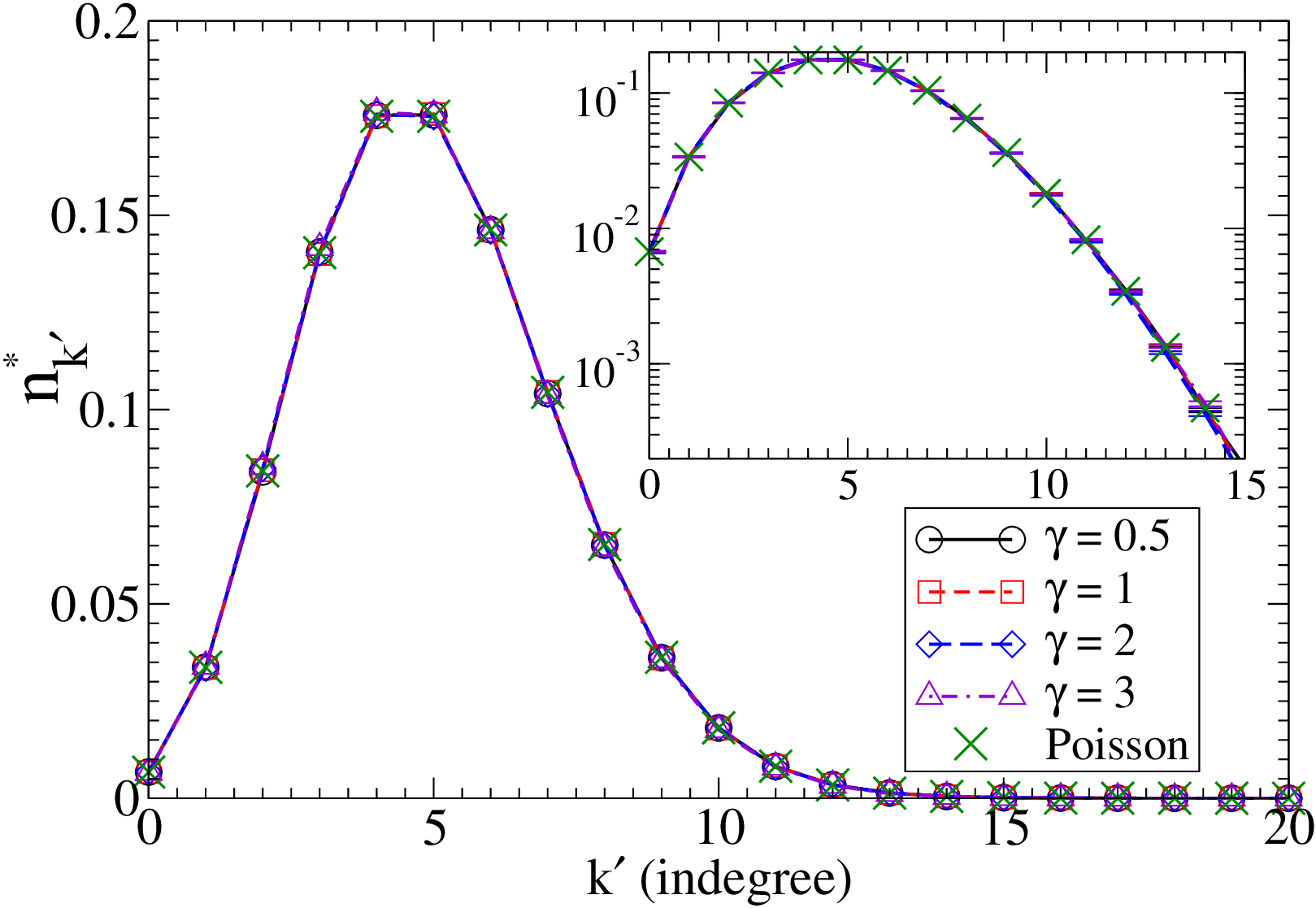} 
\caption{\label{fig:vargamma-in}Indegree distributions for $\gamma=0.5,$
1, 2, and 3 with $m=5,$ and network size $N_{0}+10^{5}$ nodes. Each
curve is averaged over fifteen runs. The distributions are identical
and they all practically overlap with the Poisson distribution, Eq.
\eqref{eq:poisson}, with $m=5.$ Inset: The same distributions shown
on a log-linear scale. The lines connecting the symbols are guides to the eye.}
\end{figure}

In contrast to the outdegree distribution, the indegree distribution
of the generalized model in the $N\gg m,k'$ limit can be described
analytically and is extremely well approximated by a Poisson distribution
with mean $m$, independent of $\gamma$ (Fig.~\ref{fig:gamma1-in}).
This can be shown as follows. Each new node makes one link per sweep
on average. When $N\gg m$, each existing node has a probability equal to $1/N$ 
of acquiring a new link per sweep on average, independent of the history
of the network. Therefore, the probability of acquiring $k'$ links
(after $m$ sweeps) for the node added at time $t$, when the total
number of nodes is $N,$ is a binomial, 
\begin{equation}
P_{k'}(N)={N \choose k'}p^{k'}(1-p)^{N-k'}
\label{eq:binom}
\end{equation}
with $p=m/N$. Thus, the final indegree distribution is an average
(ignoring the $N_{0}$ initial nodes), 
\begin{equation}
\nkin=\frac{1}{N_{\mathrm{max}}}
\sum_{N=N_{0}}^{N_{0}+N_{\mathrm{max}}}P_{k'}(N)\; .
\label{eq:nktimeaverage}
\end{equation}
In general, this cannot be calculated exactly. However, for $N\gg m,k'$,
$P_{k'}(N)$ can be approximated by a poissonian of mean $m$ \cite{RohatgiProb},
\begin{equation}
P_{k'}=\frac{m^{k'}\exp(-m)}{k'!} \;,
\label{eq:poisson}
\end{equation}
which is independent of $N$. The convergence with $N$ to this result
is fast, so that for $N_{\mathrm{max}}\gg N_{0},m,$ the sum in 
Eq.~\eqref{eq:nktimeaverage} is dominated by the $N$-independent terms.
As a result,
\begin{equation}
\nkin\approx\frac{m^{k'}\exp(-m)}{k'!}
\end{equation}
is an excellent approximation, as shown in Fig.~\ref{fig:gamma1-in}.
Computer simulations confirm the $\gamma$-independent form of the
indegree distribution, as shown in Fig.~\ref{fig:vargamma-in}.

\section{Discussion}
The mechanism that we propose and study in this paper, 
growth by preference for high indegree (number of prey) and low outdegree
(number of predators), is intended to simulate the early stages 
of the development of food webs and other transportation networks. 
To distinguish it from the more commonly studied ``rich get richer" schemes 
that produce scale-free networks, we call it 
{\it inverse preferential attachment\/}.

The outdegree distribution obtained using the generalized
form of the probability of attachment, 
$\Pi(k'_{i},k_{i})\propto(k'_{i}/k_{i})^{\gamma},$
(with $\gamma=1)$ is broader than the one obtained from the simplified
form, $\Pi(k_{i})\propto1/k_{i}$. However, both decay faster than exponentially
for large $k_{i}$. The shape of the outdegree distribution is continuously 
tunable by $\gamma$, from a exponential distribution for $\gamma=0$ to a 
delta function for $\gamma \rightarrow \infty$.
The indegree distribution does {\it not\/} depend on $\gamma$, but
only on $m$, the mean number of links per node. 
In the limit $N\gg k,m$, the indegree distribution is a poissonian. 

Some features of the networks
produced by this model, like the outdegree distribution, which decays
faster than exponentially, resemble those of some empirical and model
food webs 
\cite{Camacho:2002B,Camacho:2002A,Martinez:2000,Rossberg:2006,Rikvold:2007}.
In this, they differ sharply from the scale-free networks generated 
by the conventional ``rich get richer" preferential-attachment schemes 
\cite{Barabasi:1999,Barabasi:2002}.
However, differences from real food webs remain, such as the
correlation between the in- and outdegrees of a node. Our model produces
webs with a positive in-outdegree correlation, whereas most empirical
and model webs have a negative correlation \cite{Stouffer:2005,Rikvold:2007}. This
may be due to the unrestrained growth of our networks, which would
require an extinction process to achieve a steady state \cite{Rossberg:2006}. 
Also, this growth scheme is not designed to produce loops.
We intend to include such features in future versions of the model, 
thus enabling modeling of mature, steady-state networks.

\section*{Acknowledgments}

This research was supported by U.S.\ National Science Foundation
Grant Nos. DMR-0240078 and DMR-0444051, and by Florida State University
through the Department of Physics, the School of Computational Science,
the Center for Materials Research and Technology, and the National
High Magnetic Field Laboratory.


\end{document}